\documentclass[useAMS,usenatbib]{mn2e}

\usepackage{amsmath,amssymb,amsfonts,bm}
\usepackage{graphicx}
\usepackage{hyperref}
\usepackage{calc}
\usepackage{color}
\usepackage{wrapfig}
\usepackage{url}
\usepackage{paralist}
\usepackage{array}
\usepackage{tabularx}
\usepackage{txfonts}

\title[MassMap and Peak in CS82]{Weak lensing mass map and peak statistics in Canada-France-Hawaii Telescope 
Stripe 82 survey}

\author[Shan et al.]{HuanYuan Shan$^{1}$\thanks{E-mail: huanyuan.shan@epfl.ch}, 
Jean-Paul Kneib$^{1,2}$, 
Johan Comparat$^{2}$, 
Eric Jullo$^{2}$, 
Ald\'ee Charbonnier$^{4,5}$,
\newauthor
Thomas Erben$^{6}$,
Martin Makler$^{5}$,
Bruno Moraes$^{5}$,
Ludovic Van Waerbeke$^{7}$,
\newauthor
Fr\'ed\'eric Courbin$^{1}$,
Georges Meylan$^{1}$,
Charling Tao$^{8,9}$, 
James E. Taylor$^{10}$\\
$^{1}$Laboratoire d'astrophysique (LASTRO), Ecole Polytechnique F\'ed\'erale de Lausanne (EPFL), Observatoire de Sauverny, CH-1290 Versoix, Switzerland\\
$^{2}$Aix Marseille Universit\'e, CNRS, LAM (Laboratoire d'Astrophysique de Marseille) UMR 7326, 13388, Marseille, France\\
$^{3}$Centre de Physique des Particules de Marseille, CNRS/IN2P3-Luminy and Universit\'e de la M\'editerran\'ee, Case 907, F-13288 Marseille Cedex 9, France\\
$^{4}$Observat\'orio do Valongo, Universidade Federal do Rio de Janeiro, Ladeira do Pedro Ant\^onio 43, Sa\'ude, Rio de Janeiro, RJ 20080-090, Brazil\\
$^{5}$Centro Brasileiro de Pesquisas F\'isicas, Rua Dr. Xavier Sigaud 150, Rio de Janeiro, RJ 22290-180, Brazil\\
$^{6}$Argelander Institute for Astronomy, University of Bonn, Auf dem H{\"u}gel 71, 53121 Bonn, Germany\\
$^{7}$Department of Physics and Astronomy, University of British Columbia, 6224 Agricultural Road, Vancouver, V6T 1Z1, BC, Canada\\
$^{8}$Centre de Physique des Particules de Marseille, CNRS/IN2P3-Luminy and Aix Marseille Universit\'e de, Case 907, F-13288 Marseille Cedex 9, France\\
$^{9}$Department of Physics and Tsinghua Center for Astrophysics, Tsinghua University, Beijing, 100084, China\\
$^{10}$Department of Physics and Astronomy, University of Waterloo, 200 University Avenue West, Waterloo, Ontario, Canada N2L 3G1 }

\begin{document}

\date{Accepted \dots . Received \dots; in original form \dots}


\maketitle

\label{firstpage}

\begin{abstract}
We present a weak lensing mass map covering $\sim$124~$\rm deg^2$ of the Canada-France-Hawaii Telescope Stripe 82 Survey (CS82).
We study the statistics of rare peaks in the map, including peak abundance, the peak-peak correlation functions, and the
tangential-shear profiles around peaks. We find that the abundance of peaks detected in CS82 is consistent with predictions
from a $\Lambda$CDM cosmological model, once noise effects are properly included. The correlation functions of peaks with
different signal-to-noise ratio (SNR) are well described by power-laws, and there is a clear cross-correlation between the
SDSS-III/Constant Mass (CMASS) galaxies and high SNR peaks. The tangential shear profiles around peaks increase with peak SNR.
We fit analytical models to the tangential shear profiles, including a projected singular isothermal sphere (SIS) model
and a projected Navarro, Frenk \& White (NFW) model, plus a 2-halo term. For the high SNR peaks, the SIS model is 
rejected at $\sim$3$\sigma$. The NFW model plus a 2-halo term gives more acceptable fits to the data. Some peaks
match the positions of optically detected clusters, while others are relatively dark. Comparing dark and matched peaks,
we find a difference in lensing signal of a factor of $2$, suggesting that about half of the dark peaks are
false detection.

\end{abstract}


\section{Introduction}

Weak gravitational lensing (WL) by large scale structures (LSS) is recognized as a powerful tool for probing
the distribution of dark matter (DM) in the Universe. The details of this distribution depend both on DM particle 
properties and on the cosmological growth factor, itself a function of the equation of state, and thus weak lensing
measurements can provide important constraints on cosmology (Kneib et al.
2003; Sheldon et al. 2004; Hoekstra et al. 2004, 2005; Clowe et al. 2006; Mandelbaum et al. 2006; Rozo et al. 2010;
Leauthaud et al. 2010, 2011a, 2011b; Kneib \& Natarajan 2011).

The two-dimensional (2D) WL convergence map is proportional to the density projected along each line of sight.
High signal-to-noise ratio (SNR) peaks in the convergence map generally correspond to massive clusters
(Hamana et al. 2004). It turns out that a simple Gaussian filter of width $\theta_G\approx1~\arcmin$
is close to the optimal linear filter for cluster detection, and this choice has been extensively studied in
simulations (White et al.\ 2002; Hamana et al.\ 2004; Tang \& Fan 2005; Gavazzi \& Soucail 2007). Shape noise
from intrinsic ellipticity of galaxies and projection effects of the LSS will produce spurious noise peaks, degrading
the completeness and purity of cluster detection. Such effects can also influence the WL signals of the LSS, increasing
the SNR of smaller structures above $3\sigma$.

As a cosmological probe, the peak abundance is complementary to the WL power spectrum, and
is similar to galaxy cluster abundance (Dietrich \& Hartlap 2010; Kratochvil et al. 2010; Maturi et al. 2010;
Yang et al. 2011, 2013; Marian et al. 2012; Shan et al. 2012; Bard et al. 2013; Van Waerbeke et al. 2013).
A major advantage of WL peaks and a motivation for
their use is that they avoid the issue of having to identify genuine bound clusters and measure their masses.
Peaks can be directly compared to cosmological N-body simulations without the need to make the correspondence to 
observed or simulated ``galaxy clusters''. Since the abundance of WL peaks can be used as a cosmological tool, we expect
their clustering to also be valuable. With simulations, Marian et al. (2013) studied the high-order statistics of WL peaks,
including the stacked tangential-shear profiles and the peak-peak correlation function. They found that the
marginalized constraints are tightened by a factor of $\sim$2 compared to the peak abundance alone, the
least contributor to the error reduction being the correlation function.

First we present the WL convergence map of the $173$-tile CS82 field, and study the WL peak statistics.
We will analyze peak abundance, peak correlation functions
and the tangential shear profiles around peaks. We count the positive and negative peaks in the mass map, measure the
peak abundance as a function of SNR, and compare with the $\Lambda$ Cold Dark Matter ($\Lambda$CDM) cosmological  
model using the analytical predictions by Fan et al. (2010). We then measure higher-order statistics of
WL peaks for the first time with real data. We investigate the correlation functions of
WL peaks. For galaxies and clusters, we expect the
correlation functions of WL peaks with different SNR to be well-fitted with power laws. Furthermore, combining with
the Constant Mass galaxies (CMASS) from the Sloan Digital Sky Survey III DR10 Baryonic Oscillation Spectroscopic Survey
(SDSS-III/DR10/BOSS, Eisenstein et al. 2011; Dawson et al. 2013) experiment, we study the cross-correlation between the
CMASS galaxies and WL peaks. We also compare our WL peak detections with catalogs of overdensities detected
via the red sequence Matched-filter Probabilistic Percolation (redMaPPer) algorithm (Rykoff et al. 2013).
We fit the tangential-shear profiles of different SNR WL peaks and ``dark clumps''
(WL peaks without any obvious optical cluster counterpart) with singular isothermal sphere (SIS) profile and
Navarro-Frenk-White (NFW, Navarro, Frenk, \& White 1996) profile plus a 2-halo term.

This paper is organized in the following way. In Section~2, we describe the CS82 data used. In Section~3,
we reconstruct the 2D lensing convergence ``mass map'', and extract a catalog of peaks. In Section~4, we study the
peak statistics with peak abundance, correlation functions and tangential-shear profiles. Section~5
summarizes and discusses the results.

Throughout this paper, we adopt a fiducial, flat $\Lambda$CDM cosmological model with $\Omega_{\rm CDM}=0.226$, 
$\Omega_b=0.0455$, $\Omega_{\Lambda}=0.7285$, $\sigma_8=0.81$, $n_{\rm initial}=0.966$,
$H_0=100~h~\mathrm{km}~{\mathrm s}^{-1}~\mathrm{Mpc}^{-1}$ with $h=0.71$.

\section{CFHT/MegaCam Stripe-82 Survey and weak lensing catalog} \label{sec:CS82}

SDSS equatorial Stripe 82, which covers more than $200$ square degrees, has a high density of spectroscopic redshifts,
with $> 100,000$ redshift measurements. On-going surveys such as the SDSS-III Baryon Oscillation Spectroscopic Survey
(BOSS) and Wiggle-Z are now adding more than $> 40,000$ new spectra to this legacy.

\begin{figure}
\begin{center}
\includegraphics[width=1.0\columnwidth]{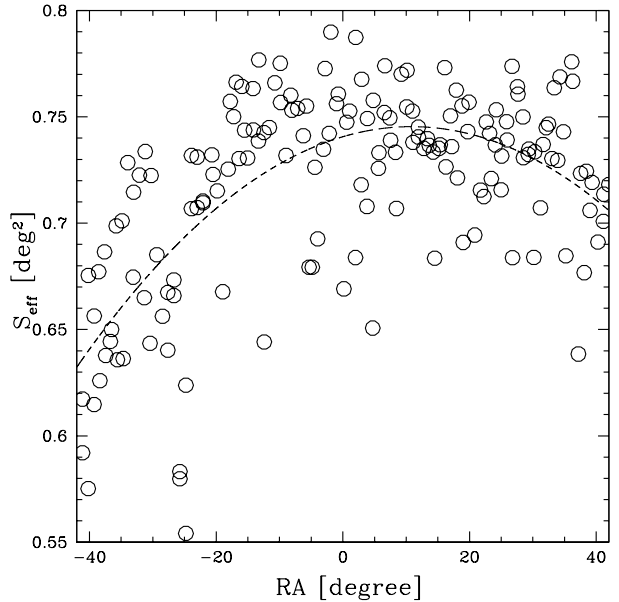}
\caption{The relation between RA and effective sky coverage. Each dot
corresponds to one CS82 tile. The dashed line shows a polynomial fit of
the data, which displays the impact of stellar density contamination.
\label{fig:ra_area}}
\end{center}
\end{figure}

The CFHT/MegaCam Stripe 82 Survey (CS82) is a large collaborative $i$-band survey between the Canadian/French
and Brazilian communities, which has been successfully conducted down to $i_{AB}=24.0$ in excellent seeing conditions
(between $0.4$ and $0.8$~arcsec with a median of $0.59$~arcsec) (Erben et al. 2014, in preparation). This area contains a
total of $173$ tiles ($165$ tiles CS82 and $8$ CFHT-LS Wide tiles). Each CS82 tile was obtained in four dithered
observations with an exposure time of $410$s, each resulting in a $5$-$\sigma$ limiting magnitude in a 2~arcsec
diameter aperture of about $i_{AB}=24.0$. After applying all the masks across the entire survey, the final effective
sky coverage drops from $173~{\rm deg}^2$ to $\sim$124~${\rm deg}^2$. Figure~\ref{fig:ra_area} shows a clear correlation
between effective sky coverage $S_{\rm eff}$ and RA direction of the data. On the two edges of stripe data, the mask
region is larger because of the higher stellar density.

The shapes of faint galaxies are measured with the Lensfit method (Miller et al 2007, 2013), the details of the calibration
and systematic effects are shown and discussed in Heymans et al. (2012). We use all galaxies with
magnitudes $i_{AB} <23.5$, signal-to-noise $\nu>10$, weight $w>0$ and FITCLASS$=0$, in which $w$ represents an inverse
variance weight accorded to each source galaxy by LensFit, and FITSCLASS is a star/galaxy classification provided by
Lensfit. The magnitude cut is quite conservative as the limiting magnitude of each tile is higher than $23.5$. These
criteria result in a total of $2,846,452$ source galaxies, and the average source surface number density is
$6.4$ galaxies per ${\rm  arcmin}^2$.

\section{Convergence map}\label{sec:mass2d}

\begin{figure*}
\begin{center}
\includegraphics[width=1.0\columnwidth]{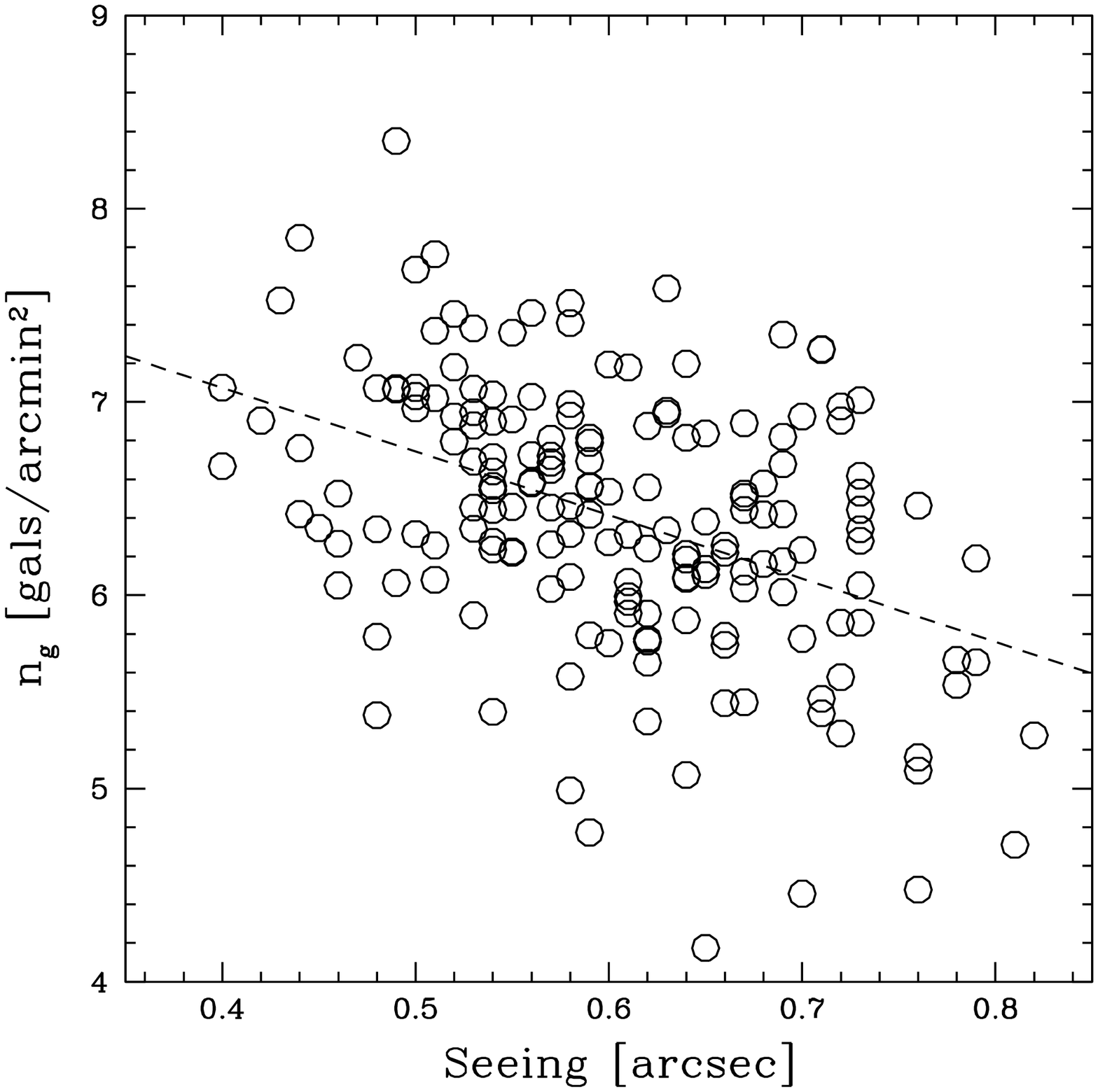}
\includegraphics[width=1.0\columnwidth]{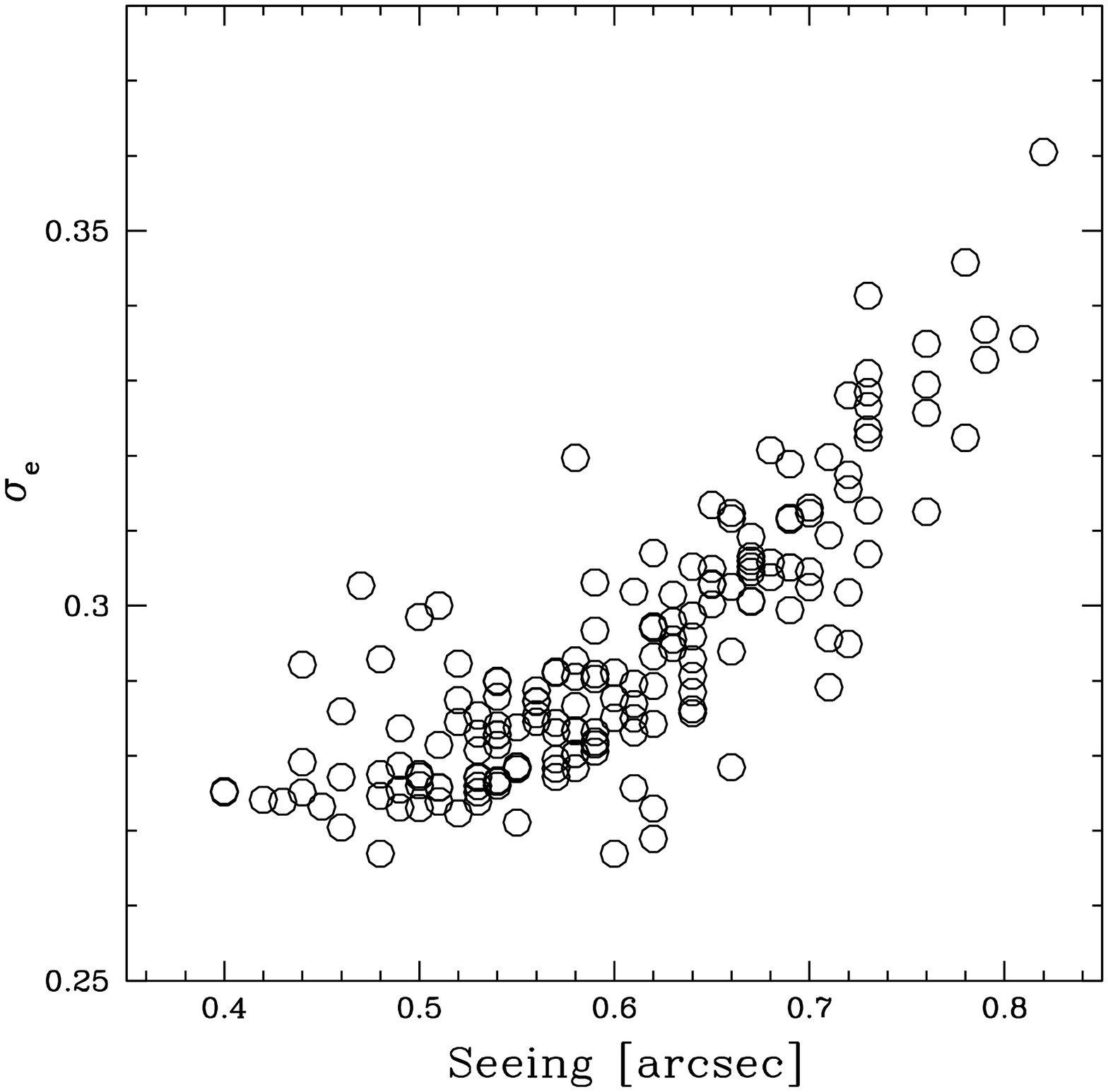}
\caption{The relation between seeing and effective galaxy number density (left panel)/intrinsic ellipticity dispersion
(right panel) of each tile. Each dot corresponds to one CS82 tile. The dashed line shows a linear fit to the data.
\label{fig:seeing_ng}}
\end{center}
\end{figure*}

The convergence field, $\kappa$, is estimated from the shear field $\gamma$ using the Kaiser \& Squires (1993)
inversion algorithm as
\begin{equation}
\kappa(\theta)=\frac{1}{\pi}\int d^2 \theta' \Re [D^*(\theta-\theta')\gamma(\theta')],
\end{equation}
where $D(\theta)=\frac{-1}{(\theta_1-i \theta_2)^2}$ is a complex convolution kernel to obtain $\kappa$ from the
shear $\gamma$. In this paper, we do the mass reconstruction per tile. The pixel scale for the binning of
$\gamma$ is $0.0586~\arcmin$.

We treat individual CS82 tiles as ``empty fields'' with average mass properties. As the tiles are degree-scale,
the net mass sheet density on this scale should be negligeable.

For the finite density of source galaxies resolved by CFHT, the scatter of their intrinsic ellipticities
means that a raw, unsmoothed convergence map $\kappa(\theta)$ will be infinitely noisy. We smooth
the convergence map by convolving it (while still in Fourier space) with a Gaussian window function,
\begin{equation}
W_G(\theta)=\frac{1}{\pi \theta_G^2} \exp \left( -\frac{\theta^2}{\theta_G^2} \right ).
\end{equation}
where $\theta_G$ is the smoothing scale. As shown by Van Waerbeke (2000), if different galaxies' intrinsic
ellipticities are uncorrelated, the statistical properties of the resulting noise field can be described by Gaussian
random field theory (Bardeen et al.\ 1986; Bond \& Efstathiou 1987) on scales where the discreteness effect of
source galaxies can be ignored. The Gaussian field is uniquely specified by the variance of the noise, which is in
turn controlled by the number of galaxies within a smoothing aperture (Kaiser \& Squires 1993; Van Waerbeke 2000)
\begin{equation}
\sigma_{\rm noise}^2=\frac{\sigma_e^2}{2} \frac{1}{2\pi \theta_G^2 n_g},
\end{equation}
where $\sigma_e$ is the rms amplitude of the intrinsic ellipticity
distribution and $n_g$ is the density of source galaxies. In Figure~\ref{fig:seeing_ng}, we show the
relation between the seeing and effective galaxy number density/intrinsic ellipticity dispersion of each tile.

We define the signal-to-noise ratio for WL detections as
\begin{equation}
\nu\equiv\frac{\kappa}{\sigma_{\rm noise}}.
\end{equation}
To define the noise level in theoretical calculations of $\nu$, we adopt a constant effective density of galaxies
equal to the mean within our survey. For our CS82 survey, we use the mean galaxy density in each
tile, which is corrected for the masked area (the area goes from $173~{\rm deg}^2$ to $124~{\rm deg}^2$) --- but do
not consider the non-uniformity of the density within each tile due to masks or galaxy clustering.

\begin{figure*}
\begin{center}
\includegraphics[width=2.0\columnwidth]{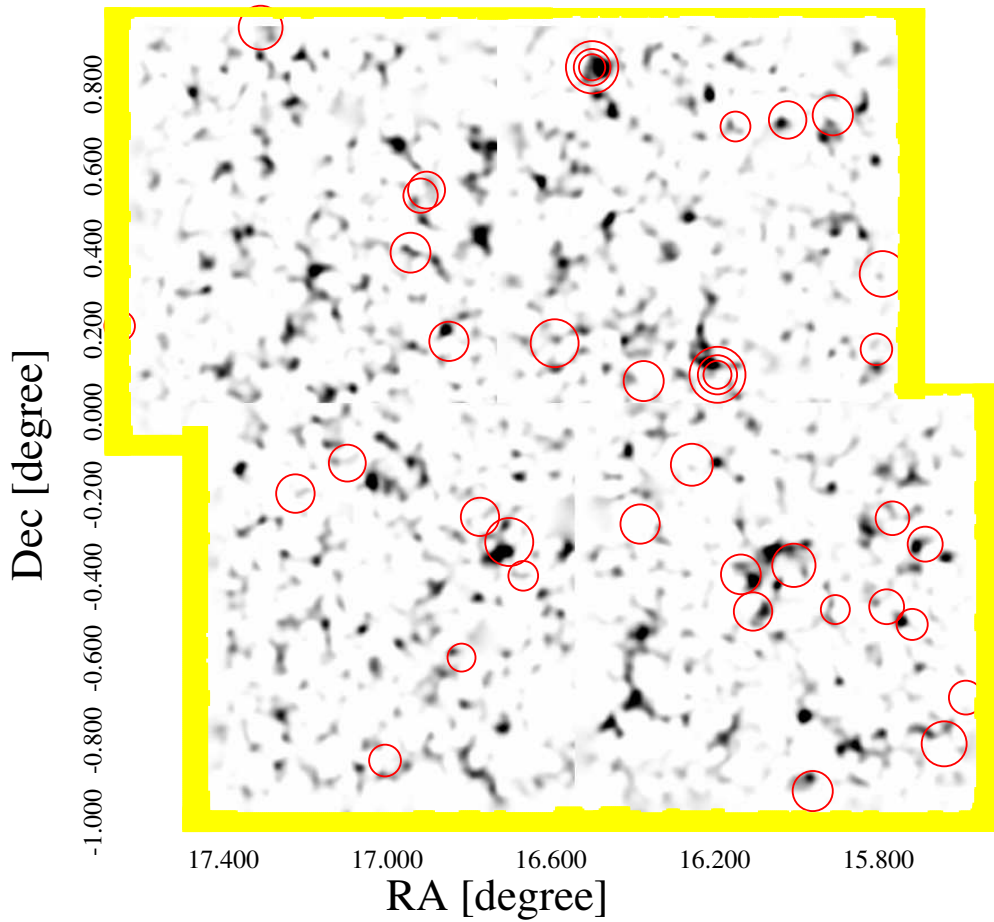}
\caption{Reconstructed ``DM mass'' convergence map for an area of
CS82 fields, smoothed with a Gaussian filter of width $\theta_G= 1'$. The red circles (multi-circles)
denote the redMaPPer (high richness) clusters. }\label{fig:map}
\end{center}
\end{figure*}

In order to better display large-scale features, we show the reconstructed ``DM mass'' convergence map for an
area of CS82 fields in Figure~\ref{fig:map} with a smoothing scale $\theta_G=1~\arcmin$.

We also compare our WL peak detections with catalogs of overdensities detected via optical observation. The red
sequence Matched-filter Probabilistic Percolation (redMaPPer) algorithm is an efficient red sequence cluster
finder developed by Rykoff et al. (2013) based on the optimized red-sequence richness estimator.
The red circles (multi-circles) denote the redMaPPer (high richness) clusters. While there may be some correlation
between our WL peaks and the positions of the redMaPPer cluster candidates, there is generally not a clear one-to-one
correspondence between individual peaks and individual clusters, presumably due to shape noise and projection effects
in the LSS.

\section{Peak statistics}

\begin{figure}
\begin{center}
\includegraphics[width=1.0\columnwidth]{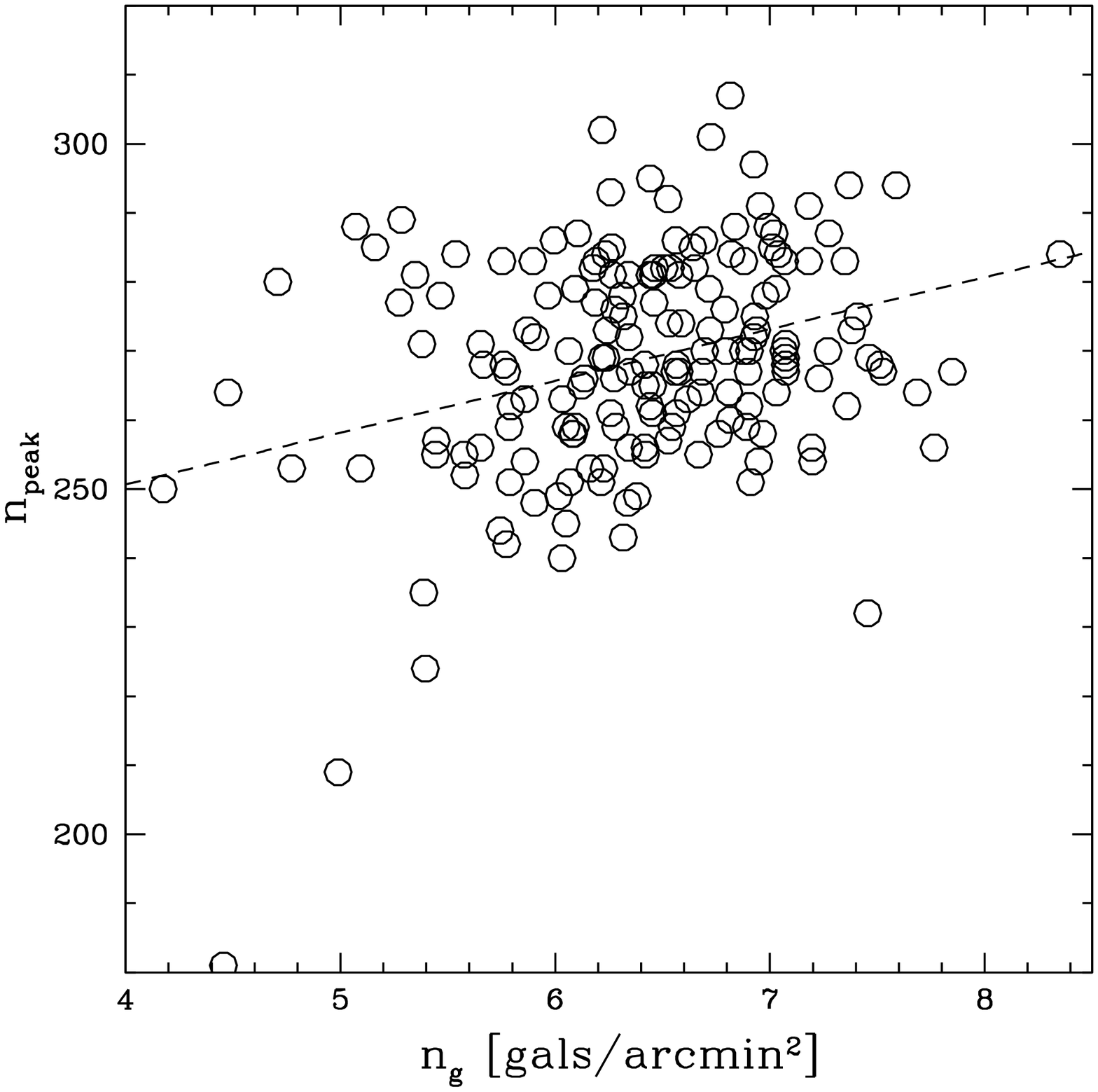}
\caption{The relation between effective galaxy number density and
the number of all WL peaks of each tile. The dashed line is a linear fit to the data.
By increasing the galaxy number density, there is an increase on the detection level,
so we can detect smaller structures as shown in this plot. However, there is a big
variance in the distribution of massive structures from field to field.
\label{fig:ng_nptot}}
\end{center}
\end{figure}

In this section, we will analyze the peaks in the WL mass maps, defining a peak as any pixel that has a higher values
of $\kappa$ than any of the surrounding eight pixels (Jain \& Van Waerbeke 2000). Considering the data, we can find that
the number of all WL peaks in each tile is related to effective galaxy number density (See Figure~\ref{fig:ng_nptot}).

We study three kinds of peak statistics: peak abundance, peak
correlation functions and the mean tangential shear around peaks based on the mass map of the CS82 survey.

\subsection{Peak abundance}\label{sec:aboud}

\begin{figure}
\begin{center}
\includegraphics[width=1.0\columnwidth]{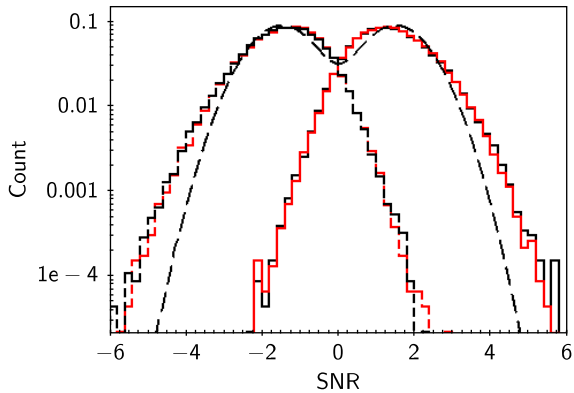}
\caption{Nomalized numbers of local maxima (solid line) and minima (dashed line) in our $E$-mode (black) and $B$-mode (red)
convergence map of the CS82 field, with smoothing scales $\theta_G=1~\arcmin$.
Local maxima can still have a slightly negative peak height if they occur along the same line of sight as a negative
noise fluctuation (or a large void), and local minima can similarly have a slightly positive peak height.
The dashed curve shows the prediction from Gaussian random field theory (van Waerbeke 2000).
\label{fig:count_peak}}
\end{center}
\end{figure}

To assess the reliability of this map, we shall first investigate the statistical properties of local maxima and minima.
Figure~\ref{fig:count_peak} shows the distribution of peak heights, as a function of SNR.
The bimodal distribution in both the $B$-mode and $E$-mode signals is dominated by positive and negative noisy
fluctuations, but an asymmetric excess in the $E$-mode signal is apparent at $\nu>3.0$.
Local minima could correspond to voids (Jain \& Van Waerbeke 2000; Miyazaki et al.\ 2002).
But the large angular extent of voids is ill-matched to our $\theta_G=1\arcmin$ filter width, and their
density contrast can never be greater than unity, so this aspect of our data is likely just noise.
The dashed curve shows the prediction from Gaussian random field theory (van Waerbeke 2000). 
The low galaxy number density of CS82 survey will introduce some Poisson noise, making even the B-mode 
peak count histograms have a non-Gaussian component.

Taking into account the effects of noise on the main-cluster-peak heights and the enhancement of the 
number of noise peaks near DM halos, Fan et al.\ (2010) developed an anlytical model incorporating the mass 
function of DM halos to calculate the statistical abundance of WL peaks over large scales. They pointed out that 
because of the mutual effects of the mass distribution of DM halos and noise, the noise peak abundance also carries 
important cosmological information, especially the information related to the density profile of DM halos. 
This model can allow us to use directly the peaks detected in the large-scale reconstructed convergence map from WL 
observations as cosmological probes without the need to differentiate true or false peaks with follow-up observations. 
We adopt the Sheth-Tormen mass function (1999) and the NFW density profile for DM halos in the calculations.

\begin{figure}
\begin{center}
\includegraphics[width=1.0\columnwidth]{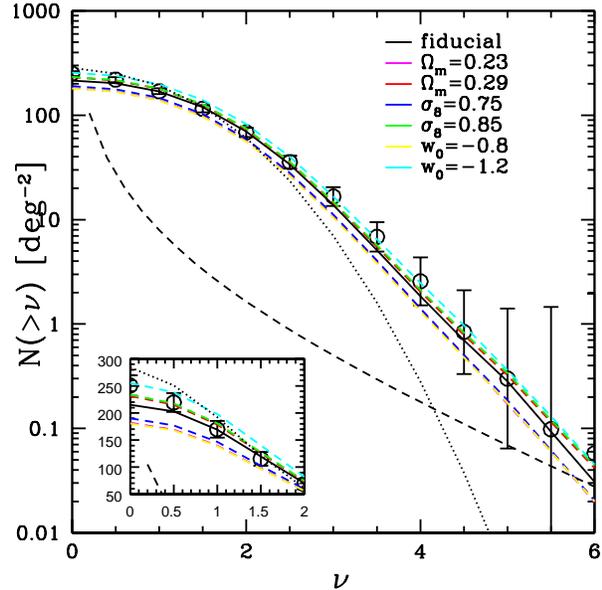}
\caption{Cumulative density of local maxima $N(>\nu)$ (black circles) using smoothing scale $\theta_G=1~\arcmin$. 
The dashed black curve show the expected peak number from true DM halos. Error bars are simply $1\sigma$
uncertainties assuming Poisson shot noise. The black solid curve shows an analytical
prediction in a fiducial $\Lambda$CDM universe (Fan et al.\ 2010), with the influence of
shape noise. The dashed color curves show different cosmological models as in Bard et al. 2013. 
The dotted curve shows the prediction from Gaussian random field theory (van Waerbeke 2000). The sub-panel shows the 
low SNR peak distribution on a linear scale.
\label{fig:count_pv}}
\end{center}
\end{figure}

Figure~\ref{fig:count_pv} recasts the peak distribution into a cumulative density of positive maxima.
The prediction from the model by Fan et al.\ (2010) is shown as solid line. In these theoretical calculations, we model
the population of background galaxies as having an intrinsic ellipticity dispersion
$\sigma_e=0.3$, density $n_g=6.4~\rm arcmin^{-2}$ and the redshift distribution $n(z)=A\frac{z^a+z^{ab}}{z^b+c}$
with $a=0.531, b=7.810, c=0.517$ and $A=0.688$. This galaxy distribution has a median redshift
$z_m=0.75$ and a mean redshift $z=0.82$ (see details in Erben et al. 2014, in prep.).
The dotted curve shows the prediction from Gaussian random field theory (van Waerbeke 2000). The measurements, especially 
the high SNR peaks, are inconsistent with a pure Gaussian noise. The model including LSS signals and shape noise are more 
reasonable.

We also show different cosmological models (dashed color curves) as in Bard et al. (2013).
At large SNR, the statistical errors are large with poor constraints on cosmological parameters.
The low SNR peak distribution is also shown on a linear scale in the sub-panel of Figure~\ref{fig:count_pv}.
The low SNR peaks contain most of the power to separate between different cosmological models, as discussed in
Kratochvil, Haiman \& May (2010). As pointed out by Yang et al. (2011), the reason why noise can boost the sensitivity
to cosmological parameters, contrary to simple intuition, is because the signal for WL peaks is a non-linear function of
the noise.

These low SNR peaks are dominated by random galaxy shape noise,
but the projection of multiple (typically, $4-8$) halos along the line of sight also contribute to the signal of the low
SNR peaks, making their number counts sensitive to cosmological parameters (Yang et al. 2011). However, the analytical
model in Fan et al. (2010) only considers the effects of shape noise. In the noise-dominated case, the distribution of peak
heights will roughly follow that expected for a Gaussian random field, but will differ in detail because of the contribution
from large-scale structures (Yang et al. 2011). A model including the projection effects of LSS should be developed before
interpreting the results in terms of cosmology.

\subsection{Auto-correlation function}\label{sec:corr}

The auto-correlation function $w(\theta)$ is measured by comparing the
actual peak distribution to a catalog of positions distributed randomly
over the unmasked region of the survey.

We use the estimator of Lancy \& Szalay (1993, LS) to calculate
$w(\theta)$, as this has been found to be the most reliable
estimator for the two-point correlation function (Kerscher et al. 2000).
The LS estimator is given by,
\begin{eqnarray}
w(\theta)&=&\frac{DD-2DR+RR}{RR}, \\
&& =1+\left ( \frac{N_{\rm rd}}{N} \right)^2 \frac{DD}{RR} - 2\left(
\frac{N_{\rm rd}}{N} \right) \frac{DR}{RR}
\end{eqnarray}
where DD, DR and RR are pair counts in bins of $\theta \pm \delta
\theta$ of the data-data, data-random and random-random points
respectively, and $N$ and $N_{\rm rd}$ are the numbers of data and random
points in the sample.

Historically, measurements of the cluster correlation function found
results consistent with a power law over scales $\rm r< 60 h^{-1}
Mpc$ or so (Bahcall \& Soneira 1983; Nichol et al. 1992; Peacock \& West 1992;
Croft et al. 1997; Gonzalez et al. 2002),
\begin{equation}
w(\theta)=\left( \frac{\theta}{\theta_0} \right)^{-\gamma},
\end{equation}
where the correlation length $\theta_0$ depends on cluster richness
(peak richness here). Thus, we fit a power-law $w(\theta)=A_w(\theta/1')^{-\gamma}$ with different SNR peaks.
A set of random points will produce $\gamma=0$.

\begin{figure}
\begin{center}
\includegraphics[angle=0.0, width=1.0\columnwidth]{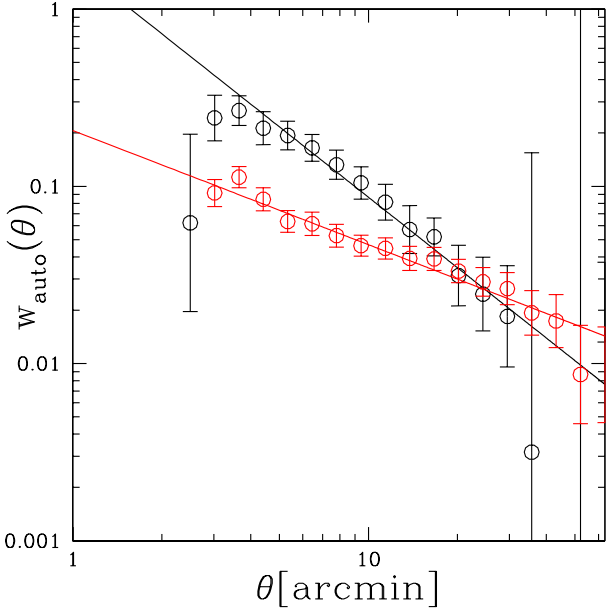}
\caption{Auto-correlation functions of WL peaks with SNR $|\nu|>0$ (red points) and $|\nu|>2$
(black points). The drop of the auto-correlation function at small scales depends on the
smoothing scale used in the analysis. \label{fig:auto}}
\end{center}
\end{figure}

In Figure~\ref{fig:auto}, we show the auto-correlation functions of WL peaks with SNR $\nu>0$ and $\nu>2$. We
fit the measured correlation function for $\theta>4~\rm arcmin$. The solid lines are the fitted power law.
The auto-correlation function of peaks can be well fitted with a power law
$w_{pp}(\theta)=A_w(\theta/1')^{-\gamma}$ (see Table~\ref{tab:tab1}). For the peaks with $\nu>0$, the
exponent of the power law has a value $\sim$0.64, which is even lower than the angular correlation
function of galaxies with $\gamma \sim$0.8 (Zehavi et al. 2002). This suggests that the $\nu>0$ peaks include lots of
small structures and also noise peaks. For the peaks with $\nu>2$ which are related to more massive structures and
less noise peaks, we find $\gamma_{\nu>2}=1.32 \pm 0.01$, which is close to the measured auto-correlation
functions of SDSS clusters with $\gamma \sim$0.8-1.3 (Estrada et al. 2009; Hong et al. 2012). The number of the
higher SNR peaks is too small to be well fitted with a power-law. Note that there is a turn-around at scale of
$\sim$3~$\rm arcmin$, which depends on the size of the Gaussian smoothing scale applied to the shear data.

\begin{table}
\centering
\caption{Slope of the power-law fit to $w(\theta)$}
\begin{tabular}{ccccc}
\hline
\hline
SNR &  $\gamma$ & $A_{w}$\\
\hline
$\nu>0$ & $0.64 \pm 0.01$ & $0.20 \pm 0.01$\\
$\nu>2$ & $1.32 \pm 0.01$ & $1.59 \pm 0.03$ \\
\hline
\end{tabular}
\label{tab:tab1}
\end{table}

\subsection{Cross-correlation function}

Because WL peaks are related to LSS in the Universe, we expect that there to be a cross-correlation between
WL peaks and biased systems, such as clusters and massive galaxies. However, the redMaPPer catalog does not contain
enough objects, $\sim$432 clusters with $0.1<z<0.6$ and richness $\lambda>20$, to estimate a correlation function.
Therefore, we use CMASS galaxies instead.

In this section, we present the cross-correlation functions between CMASS galaxies and WL peaks with
$\nu>2.0$. The CMASS sample is the SDSS-III/BOSS experiment BAO tracer (Dawson et al. 2013). The parent catalog of
CMASS selection on Stripe 82 contains $22,034$ tracers, covering $\sim$98~$\rm deg^{-2}$ of the CS82 region. As in
Comparat et al. (2013), we use the complete CMASS selection, not only the galaxies confirmed by spectroscopy, in order
to avoid fiber collision issues. The mean redshift is $0.53$ with a dispersion of $0.1$. On the same scale where 
we have determined the auto-correlation functions of WL peaks, we measure the slope of the auto-correlation function for
CMASS galaxies to be $\gamma \sim$0.73$\pm$0.01. We conclude that the WL peaks with $\nu>2.0$ are more biased than
CMASS galaxies, suggesting that these peaks are related to groups or clusters as expected.

\begin{figure}
\begin{center}
\includegraphics[angle=0.0, width=1.0\columnwidth]{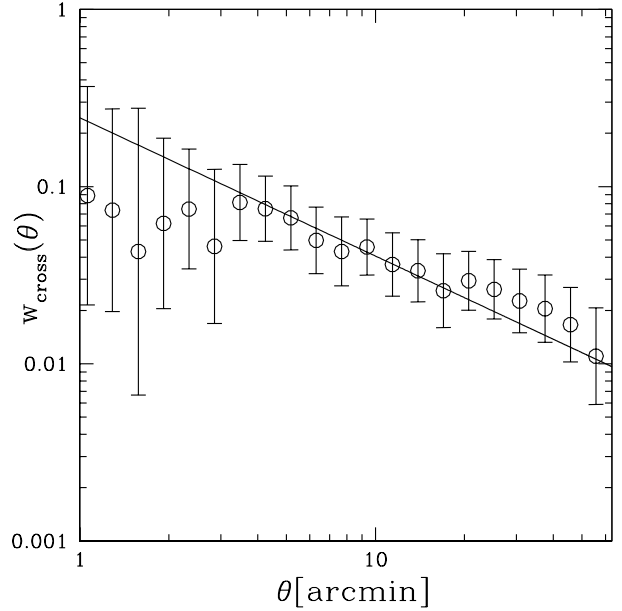}
\caption{Cross-correlation functions between CMASS galaxies and WL peaks with SNR $|\nu|>2$. \label{fig:cmass}}
\end{center}
\end{figure}

Eventually we checked the cross-correlation between CMASS galaxies and WL peaks. Figure~\ref{fig:cmass} shows that this
can also be fit with a power law $w_{cp}(\theta)=A_w(\theta/1')^{-\gamma}$. The slope of cross correlation is $0.78\pm0.01$.

\subsection{Tangential shear}\label{sec:tan}

In this section, we estimate the average tangential shear profile of WL peaks. Stacking the signal from many peaks
can reduce the contribution from shape noise, uncorrelated structures along the line of sight and
substructures. We calculate the excess surface mass density $\Delta \Sigma(R)=\Sigma_{\rm crit}g_t(R)$,
where $g_t(R)$ is the tangential shear and $\Sigma_{\rm crit}$ is the critical surface density. The mean source and lens redshifts
$<z_s>$ and $<z_l>$ are {$0.75$} and $0.45$, respectively.

\subsubsection{Matched redMaPPer peaks}

Tangential shear measurements require the identification of the DM density peak, but WL peaks
are expected to be offset of the centers of the main DM halos associated with them (Yang et al. 2013).
In this paper, halo centers are assumed to contain central galaxies (CGs) which can be used as good tracers, on condition
that they can be correctly identified.

\begin{figure}
\begin{center}
\includegraphics[width=1.0\columnwidth]{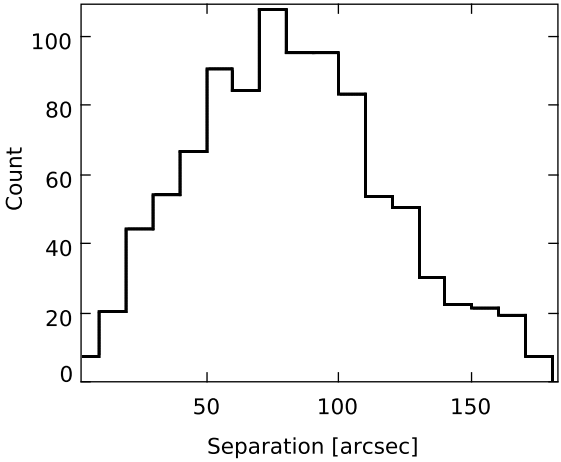}
\caption{The separation histogram of redMaPPer matched peaks.
\label{fig:matchedpeak_d}}
\end{center}
\end{figure}

As in Shan et al. (2012), we search for matched redMaPPer clusters within a $3.0\arcmin$ radius of peaks that appear in
the WL mass map. This search radius is chosen to be larger than the smoothing scale, but smaller than the angular virial
radius of a massive cluster at $0.1<z<0.9$ (Hamana et al.\ 2004). If more than one pair exists within $3.0\arcmin$, we
adopt the closest match as the primary candidate. In total, $19$ redMaPPer clusters have no corresponding WL peaks. We show
the separation histogram of redMaPPer matched peaks in Figure~\ref{fig:matchedpeak_d}. The separation of WL peaks and
optical centers is from various systematic noise sources, such as the effect of projected LSS
(Gavazzi \& Scoucail 2007; Geller et al. 2010), smoothing of the mass maps, and shape noise in the maps (Dietrich et al. 2012).
The purity, defined as the fraction of peaks above a given detection threshold
$\nu_{\rm th}$ that are associated with an optically detected cluster (Shan et al. 2012) of our sample is much lower
($\sim$15$\%$ for $\nu>3.5$ WL peaks), which is due to the lower galaxy number density.

\begin{figure}
\begin{center}
\includegraphics[angle=0.0, width=1.0\columnwidth]{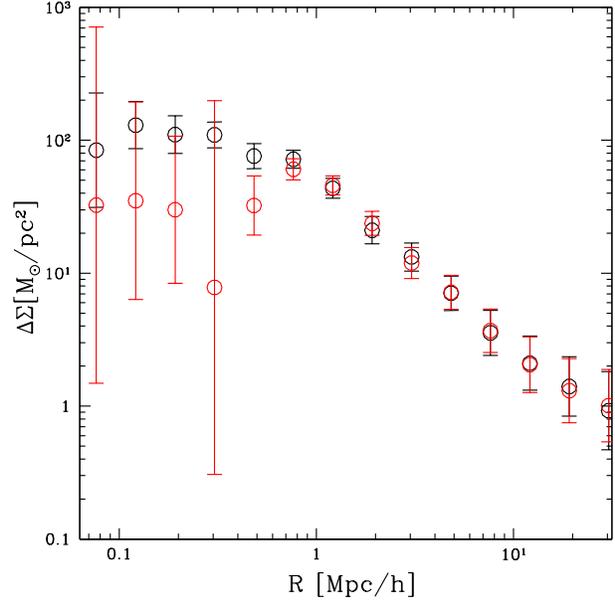}
\caption{The measured excess surface mass density $\Delta \Sigma$, obtained by
stacking the tangential shear signals for the redMaPPer matched peaks with $\nu>3.0$ using the WL
peak positions (red points) and the matched redMaPPer cluster center positions (black points), respectively.
\label{fig:center}}
\end{center}
\end{figure}

In Figure~\ref{fig:center}, we present $\Delta \Sigma$ of redMaPPer matched peaks with SNR $\nu>3$
using the WL peak positions (red points) and the matched redMaPPer cluster center positions (black points),
respectively. The WL peak positions can introduce noise on the tangential shear measurement. As in Johnston
et al. (2007), Leauthaud et al. (2010) and George et al. (2012), centroid errors will lead to a smoothing of the
lensing signal on small scales and to an underestimate of halo masses. On large scales ($R>0.7~\rm Mpc$), the
tangential shear signals are almost the same. Thus, we will use the matched redMaPPer cluster centers in this paper.

\begin{figure}
\begin{center}
\includegraphics[angle=0.0, width=1.0\columnwidth]{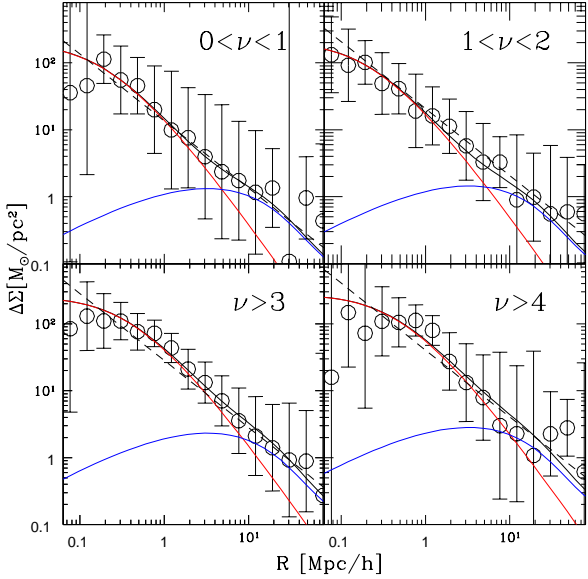}
\caption{{Black circles show the measured excess surface mass density $\Delta \Sigma$ for $4$ samples of WL peaks 
with different SNR. The solid and dashed black curves are the best-fit NFW model plus a 2-halo term
and SIS model, respectively. The blue and red curves are main galaxy cluster halo and 2-halo term, respectively.
The SIS model is more disfavored. The corresponding best-fit parameters are listed in Table~\ref{tab:tab2}. }
\label{fig:peak}}
\end{center}
\end{figure}

In Figure~\ref{fig:peak}, we present $\Delta \Sigma$ of matched WL peaks with SNR $0<\nu<1$ (top-left panel),
$1<\nu<2$ (top-right panel), $\nu>3$ (bottom-left panel) and $\nu>4$ (bottom-right panel). From the figure,
the higher SNR peaks have higher $\Delta \Sigma$: there is about an order of magnitude difference between the
lensing signals corresponding to the $\nu>4$ and $0<\nu<1$ bins.

Low SNR peaks often include contributions from several halos or structures along the line of sight, thus
they are relatively insensitive to the inner profile of individual halos (Yang et al. 2013), but reflect instead the
``2-halo term'', or halo-halo clustering. We fit the stacked profiles with both a SIS and a NFW model plus a 2-halo term
following Covone et al. (2014). Fitting results are given in Table~\ref{tab:tab2}. Note that although the full NFW model has
two free parameters, we assume the Bullock et al. (2001) relation between concentration c and virial mass $M_{\rm vir}$ seen
in numerical simulations, leaving only a single parameter to fit.

\begin{table*}
\centering
\caption{Density Profile Models in Figure~\ref{fig:peak}.}
\begin{tabular}{ccccc}
\hline
\hline
SNR &  $M_{\rm vir}$ & $\chi^2_{\rm NFW}/\rm d.o.f$ & $\sigma_v$ & $\chi^2_{\rm SIS}/\rm d.o.f$ \\
    &  $10^{14} M_{\odot}/h$  &  & km/s  &  \\
\hline
$0<\nu<1$ & $0.97^{+0.26}_{-0.25}$ & $0.47$ & $536.65^{+144.45}_{-152.34}$ & $2.38$ \\
$1<\nu<2$ & $1.25^{+0.31}_{-0.28}$ & $0.43$ & $554.53^{+135.67}_{-87.17}$ & $1.54$ \\
$\nu>3$ & $4.07^{+0.98}_{-0.83}$ & $0.76$ & $775.41^{+129.91}_{-92.79}$ & $4.54$ \\
$\nu>4$ & $6.28^{+1.36}_{-1.21}$ & $1.27$ & $928.37^{+150.94}_{-220.32}$ & $3.27$ \\
\hline
\end{tabular}
\label{tab:tab2}
\end{table*}

As expected, the mass of NFW profiles and the velocity dispersion of SIS profiles increase with SNR. The SIS model is
strongly disfavored for the high SNR peaks $\nu>3$ and $\nu>4$, and is rejected at the $3~\sigma$ level.  This could be
due to the inner slope of the DM mass density in halos (Rocha et al. 2013). The NFW model plus a 2-halo term gives
more acceptable fits to the data.

\subsubsection{Dark clump peaks}

\begin{figure}
\begin{center}
\includegraphics[angle=0.0, width=1.0\columnwidth]{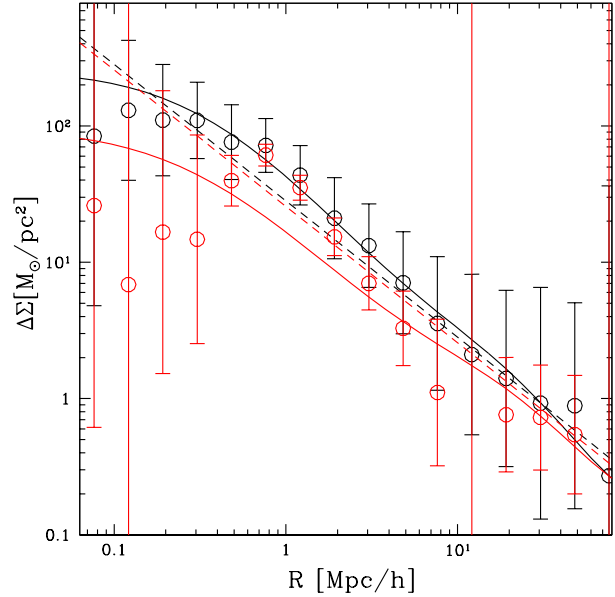}
\caption{The measured $\Delta \Sigma$ for the $\nu>3.0$ peaks, which is
obtained by stacking the distortion signals for the peaks with the matched redMaPPer
cluster center positions (black) and without any matched clusters (red), respectively. The solid and dashed
lines are best-fitted NFW and SIS models, respectively, together with a 2-halo term.
\label{fig:dark}}
\end{center}
\end{figure}

We also study the profiles of the peaks without any matched clusters (``dark clumps'' hereafter), which could include
both smaller structures and noisy peaks. The same redshift distribution with $z_l=0.45$ as the redMaPPer matched peaks
is used for ``dark clumps''.
Because of mis-centering problems, only $\Delta \Sigma$ on large scales $\rm R>0.7~\rm Mpc$
can be used. For the peaks with $\nu>3.0$, the tangential shear signals of dark clumps on large scales are lower
than the redMaPPer matched peaks (see Figure~\ref{fig:dark}). The best fits of the dark clumps are:
$M_{\rm vir}=2.40^{+1.57}_{-1.23} \times 10^{14} M_{\odot}/h$ with $\chi^2=13.80$
(NFW profile plus 2-halo term) and $\sigma_v=740.86^{+531.37}_{-518.25} \rm km/s$ with
$\chi^2=22.17$ (SIS profile). The SIS profile is even more strongly disfavored for dark clumps. It is rejected
at $22\sigma$. Comparing with the matched clumps (to the redMaPPer clusters) in Table~\ref{tab:tab2}, there is a
difference in virial mass of a factor of $2$.

\section{Conclusions}\label{sec:conc}

In this paper, we have reconstructed the WL convergence map of CS82 fields. With the peaks in the WL mass map, we 
further study three kinds of WL peak statistics, the peak height distribution, the peak auto-correlation function,
and the mean tangential shear around peaks. This is the first measurement of high-order statistics of
WL peaks with real data.

The use of peak abundance as a cosmological tool have been discussed extensively in the literature (Dietrich \& Hartlap 2010;
Kratochvil et al. 2010; Yang et al. 2011, 2013;  Marian et al. 2012, 2013; Shan et al. 2012; Bard et al. 2013). In our paper, we
measure the abundance of peaks as a function of SNR, and compare it with the analytical prediction in Fan et al.
(2010). The peak abundance detected in CS82 is consistent with predictions from a $\Lambda$CDM cosmological
model, once shape noise effects are properly included. If other noise effects, including projection effects and
mask effects, were included accurately in analytic models, we suggest that WL peak abundance could become
a better method to constrain cosmology than pure cluster counts, because we could use the information contained
in the large number of low SNR peaks.

The slope and amplitude of the peak auto-correlation functions depends on the SNR of WL peaks. The auto-correlation
function of different SNR WL peaks $\nu>0$ and $\nu>2$ can be well fitted with power laws with the following slopes:
$\gamma_{\nu>0}=0.64\pm 0.01$ and $\gamma_{\nu>2}=1.32\pm 0.01$. {We conclude that the WL peaks with $\nu>2.0$ are more
biased than CMASS galaxies, suggesting that these peaks are related to groups or clusters as expected.} Combining with the
CMASS galaxies, the cross-correlation with a power-law slope $\gamma \sim$0.78 between CMASS galaxies and high SNR peaks can be
found.

We also fit spherical models to the mean tangential shear profiles around peaks, including the singular isothermal sphere (SIS)
model and Navarro, Frenk \& White (NFW) model plus 2-halo term. The SIS model is strongly disfavored for the high SNR
peaks $\nu>3$ and $\nu>4$, which is rejected at $3~\sigma$. The NFW model plus 2-halo term gives more
acceptable fits to the data. We also compared the dark and matched clumps (to the redMaPPer clusters) and found that there
is a difference in virial mass of a factor of $2$, assuming the matched and unmatched peaks have the
same mass function. This could indicate that approximately half of the dark clumps are false detections, in the sense
that they do not correspond to single massive halos along the line of sight. This assumption would require better data
(such as extensive spectroscopic follow-up) to validate, however.

The high SNR peaks in the WL mass map are related to the LSS in the Universe. In an upcoming paper,
we will constrain cosmology with WL peak statistics. Future surveys, such as the Dark Energy Survey (DES), 
Large Synopic Survey Telescope (LSST), Kunlun Dark Universe Survey Telescope (KDUST) and Euclid
surveys, will allow us to map WL peaks throughout much larger cosmological volumes, thus probing cosmology more sensitively.

\chapter{\flushright{\bf{Acknowledgments}}}
\flushleft{Based on observations obtained with MegaPrime/MegaCam, a joint project of CFHT and CEA/DAPNIA, at the
Canada-France-Hawaii Telescope (CFHT), which is operated by the National Research Council (NRC) of Canada,
the Institut National des Science de l'Univers of the Centre National de la Recherche Scientifique (CNRS) of
France, and the University of Hawaii. The Brazilian partnership on CFHT is managed by the Laborat\'{o}rio Nacional
de Astrof\'isica (LNA). This work made use of the CHE cluster, managed and funded by ICRA/CBPF/MCTI, with financial
support from FINEP and FAPERJ. We thank the support of the Laborat\'{o}rio Interinstitucional de e-Astronomia (LIneA).
We thank the CFHTLenS team.

The authors thank Zoltan Haiman for useful discussions. This research was supported by a Marie Curie International Incoming
Fellowship within the $7^{th}$ European Community Framework Programme. HYS acknowledges the support from Swiss National Science
Foundation (SNSF) and NSFC of China under grants 11103011. JPK acknowledges support from the ERC advanced grand LIDA. TE is
supported by the Deutsche Forschungsgemeinschaft through project ER 327/3-1 and by the
Transregional Collaborative Research Centre TR 33 - `The Dark Universe'.}

\end{document}